PLEASE CUT HERE-----------------------------------
\documentstyle[preprint,revtex]{aps}
\begin{document}
\draft
\begin{title}
Anyonic states in Chern-Simons theory
\end{title}
\author{Kurt Haller and Edwin Lim-Lombridas}
\begin{instit}
Department of Physics, University of Connecticut, Storrs,
Connecticut 06269
\end{instit}
\begin{abstract}
We discuss the canonical quantization of Chern-Simons theory
in $2+1$ dimensions, minimally coupled to a Dirac spinor
field. Gauss's law and the gauge condition, $A_0 = 0$, are
implemented by embedding the formulation in an appropriate
physical subspace. We find two kinds of charged particle
states in this model. One kind has a rotational anomaly in
the form of arbitrary phases that develop in $2\pi$
rotations; the other kind rotates ``normally''---i.e.,
charged states only change sign in $2\pi$ rotations. The
rotational anomaly has nothing to do with the implementation
of Gauss's law. It is possible to inadvertently produce
these anomalous states in the process of implementing
Gauss's law, but it is also possible to implement Gauss's
law without producing rotational anomalies. Moreover, states
with or without rotational anomalies obey ordinary Fermi
statistics.
\end{abstract}

\narrowtext

In recent work, attention has been directed to the question
of how, and indeed whether, $(2+1)$-dimensional gauge
theories develop anyons---i.e., particle states that have
properties characteristic of neither fermions nor bosons.
Examination of the literature reveals a lack of unanimity on
this question. Some authors report finding anyons in
$(2+1)$-dimensional gauge theories, \cite{semenoff,luscher}
others question these claims. \cite{hagen,boyanovsky,ys} In
some work, attention is focused on anomalies in the angular
momentum operator. \cite{swanson} In other work it is
claimed that graded algebras develop among the gauge-
invariant operators that correspond to the charged states
that implement Gauss's law (or at least its long-range
component). \cite{semenoff,luscher} In earlier work, we
studied the topologically massive Maxwell-Chern-Simons (MCS)
theory and found that, in a canonically quantized theory in
which Gauss's law and the temporal gauge ($A_0=0$) are
implemented, the canonical angular momentum rotates charged
states without anomalies, so that the state vector for an
electron $|e\rangle$, returns to $-|e\rangle$ after a $2\pi$
rotation. \cite{hl} We also demonstrated that ``normal''
anticommutation rules govern the gauge-invariant operators
that project, from the vacuum, charged fermions which obey
Gauss's law. Moreover, in our work, these gauge-invariant
operators arise naturally within the formalism, and do not
need to be constructed {\it ad hoc\/}. Electrons in MCS
theory therefore are ordinary and unexceptional fermions,
albeit in $2+1$ dimensions.

In this work, we describe an investigation of Chern-Simons
(CS) theory, in which the CS term is the only kinetic energy
term, but the gauge field is still minimally coupled to a
charged fermion field. As has been noted, such theories do
not possess any observable propagating modes of the gauge
field. \cite{deser} We treat this model much as we have
previously treated the topologically massive MCS theory.
\cite{hl} We introduce a gauge-fixing field in such a way
that $A_0$ has a conjugate momentum and obeys canonical
commutation rules. Although, as in our treatment of MCS
theory, Gauss's law and the gauge condition are not primary
constraints, there nevertheless are other primary
constraints in CS theory. These primary constraints relate
the canonically conjugate momentum of $A_1$ to $A_2$, and
vice versa, so that the gauge field $A_i$ will be subject to
Dirac rather than Poisson commutation rules. Furthermore,
all components of the CS gauge field, $A_1$ and $A_2$ as
well as $A_0$, must be represented entirely in terms of
ghost operators, which can mediate interactions between
charges and currents but do not carry energy-momentum, and
have no probability of being observed. Neither longitudinal
nor transverse components of the CS fields have any
propagating particle-like excitations.

The Lagrangian for this model is given by
\begin{eqnarray}
{\cal L} &=& \case 1/4 m\epsilon_{ln}(F_{ln}A_0 -
2F_{n0}A_l) - \partial_0A_0G\nonumber\\
&+& j_lA_l - j_0A_0 + \bar{\psi}(i\gamma^\mu \partial_\mu -
M)\psi
\label{eq:L}
\end{eqnarray}
where $F_{ln} = \partial_nA_l - \partial_lA_n$ and $F_{l0} =
\partial_lA_0 + \partial_0A_l$. We follow conventions
identical to those in Ref.~\cite{hl}.

The Euler-Lagrange equations are
\begin{equation}
m\epsilon_{ij}F_{j0} - j_i = 0,
\label{eq:ampere1}
\end{equation}
\begin{equation}
\case 1/2 m\epsilon_{ij}F_{ij} + \partial_0G - j_0 = 0,
\label{eq:gaussmotion}
\end{equation}
\begin{equation}
\partial_0A_0 = 0,
\label{eq:DoAo}
\end{equation}
and
\begin{equation}
(M - i\gamma^\mu D_\mu)\psi = 0,
\label{eq:diraceq}
\end{equation}
where $D_\mu$ is the gauge-covariant derivative $D_{\mu} =
\partial_{\mu} + ieA_{\mu}$. Current conservation leads to
\begin{equation}
\partial_0\partial_0G = 0.
\label{eq:DoDoG}
\end{equation}

The momenta conjugate to the fields are given by
\begin{equation}
\Pi_0 = -G,
\label{eq:Pi0}
\end{equation}
and
\begin{equation}
\Pi_i = \case 1/2 m\epsilon_{ij}A_j.
\label{eq:Pii}
\end{equation}
The Hamiltonian density is given by
\begin{equation}
{\cal H} = -\case 1/2 m\epsilon_{ij}F_{ij}A_0 + j_0A_0 -
j_iA_i + {\cal H}_{e\bar{e}}
\label{eq:hdensity}
\end{equation}
where ${\cal H}_{e\bar{e}} = \psi^\dagger(\gamma^0M -
i\gamma^0\gamma^n\partial_n)\psi$ and the total derivative
$\partial_j(\case 1/2 m\epsilon_{ij}A_iA_0)$ has been
dropped.

The equal-time commutation (and anticommutation) rules
(ETCR) are
\begin{equation}
[A_0({\bf x}), G({\bf y})] = -i\delta({\bf x - y}),
\label{eq:AoG}
\end{equation}
\begin{equation}
[A_i({\bf x}), A_j({\bf y})] = (i/m)\epsilon_{ij}\delta({\bf
x - y}),
\label{eq:AiAj}
\end{equation}
and
\begin{equation}
\{\psi_\alpha({\bf x}), \psi_\beta{}^\dagger({\bf y})\} =
\delta_{\alpha,\beta}\delta({\bf x - y}),
\label{eq:psipsi2}
\end{equation}
where Eq.~(\ref{eq:AiAj}) is the Dirac rather than the
Poisson commutation rule, and represents the influence of
the constraint given in Eq.~(\ref{eq:Pii}).
\cite{dirac,faddeev} We now construct the following momentum
space expansions of the gauge fields in such a way that the
ETCR given in Eqs.~(\ref{eq:AoG}) and (\ref{eq:AiAj}) are
satisfied (all summations are over ${\bf k}$):
\begin{eqnarray}
A_i({\bf x}) &=& ({2m^{3/2}})^{-1}\sum{} k_i[a_R({\bf k}) -
a^\star_R(-{\bf k})]e^{i{\bf k \cdot x}}\nonumber\\
&+& i\sqrt{m}\sum{}\frac{\epsilon_{ij}k_j}{k^2}[a_Q({\bf k})
+ a^\star_Q(-{\bf k})]e^{i{\bf k \cdot x}}\nonumber\\
&+& i\sum{}\phi({\bf k})k_i[a_Q({\bf k}) + a^\star_Q(-{\bf
k})]e^{i{\bf k \cdot x}},
\label{eq:Ai}
\end{eqnarray}
\begin{equation}
A_0({\bf x}) = im^{-1/2}\sum{}[a_Q({\bf k}) - a^\star_Q(-
{\bf k})]e^{i{\bf k \cdot x}},
\label{eq:Ao}
\end{equation}
and
\begin{equation}
G({\bf x}) = -\case 1/2 \sqrt{m}\sum{}[a_R({\bf k}) +
a^\star_R(-{\bf k})]e^{i{\bf k \cdot x}}
\label{eq:G}
\end{equation}
where $\phi({\bf k})$ is some arbitrary real and even
function of ${\bf k}$. The explicit form of $\phi({\bf k})$
is immaterial to the commutation rules given in
Eqs.~(\ref{eq:AoG}) and (\ref{eq:AiAj}); its form as well as
its inclusion in Eq.~(\ref{eq:Ai}) are therefore entirely
optional. The operators $a_Q({\bf k})$ and $a_R({\bf k})$
and their Hermitian adjoints $a^\star_Q({\bf k})$ and
$a^\star_R({\bf k})$ are the same ghost operators previously
used for the MCS theory; \cite{hl} they obey the commutation
rules
\begin{equation}
[a_Q({\bf k}), a^\star_R({\bf k})] = [a_R({\bf k}),
a^\star_Q({\bf q})] = \delta_{\bf k, q},
\end{equation}
and
\begin{equation}
[a_Q({\bf k}), a^\star_Q({\bf q})] = [a_R({\bf k}),
a^\star_R({\bf q})] = 0.
\end{equation}
The Hamiltonian $H =\int d{\bf x}\,{\cal H}({\bf x}) =  H_0
+ H_{\mbox{\scriptsize\rm int}}$, where $H_0$ and
$H_{\mbox{\scriptsize\rm int}}$ are given by
\FL
\begin{eqnarray}
H_0 &=& -\int{}d\,{\bf x}\ \case 1/2 m\epsilon_{ij}F_{ij}A_0
+ H_{e\bar{e}}\nonumber\\
&=& im\sum{}[a_Q({\bf k})a_Q(-{\bf k}) - a_Q^\star({\bf
k})a_Q^\star(-{\bf k})] + H_{e\bar{e}}
\label{eq:hop}
\end{eqnarray}
with $H_{e\bar{e}} = \int d{\bf x}\,{\cal H}_{e\bar{e}}({\bf
x})$ and
\begin{eqnarray}
H_{\mbox{\scriptsize\rm int}} &=& im^{-1/2}\sum{}[a_Q({\bf
k})j_0(-{\bf k}) - a_Q^\star({\bf k})j_0({\bf
k})]\nonumber\\
&-&({2m^{3/2}})^{-1}\sum{}k_i[a_R({\bf k})j_i(-{\bf k}) +
a_R^\star({\bf k})j_i({\bf k})]\nonumber\\
&-& i\sqrt{m}\sum{}\frac{\epsilon_{ij}k_j}{k^2}[a_Q({\bf
k})j_i(-{\bf k}) - a_Q^\star({\bf k})j_i({\bf
k})]\nonumber\\
&-& i\sum{}\phi({\bf k})k_i[a_Q({\bf k})j_i(-{\bf k}) -
a_Q^\star({\bf k})j_i({\bf k})].
\label{eq:hip}
\end{eqnarray}
$H_0$ and $H_{\mbox{\scriptsize\rm int}}$ operate in a
Hilbert space $\{|h\rangle\}$ that very closely resembles
the one used in Ref.~\cite{hl}; $\{|h\rangle\}$ is based on
the perturbative vacuum $|0\rangle$ annihilated by all
annihilation operators, $a_Q({\bf k})$ and $a_R({\bf k})$ as
well as the electron and positron annihilation operators
$e({\bf k})$ and $\bar{e}({\bf k})$, respectively. The
Hilbert space $\{|h\rangle\}$ contains a subspace
$\{|n\rangle\}$ that consists of all multiparticle
electron-positron states of the form $|N\rangle =
\bar{e}^\dagger({\bf q}_1)\cdots \bar{e}^\dagger({\bf
q}_l){e}^\dagger({\bf p}_1)\cdots{e}^\dagger({\bf
p}_n)|0\rangle$, as well as all other states of the form
$a^\star_Q({\bf k}_1)\cdots a_Q^\star({\bf k}_i)|N\rangle$.
$H_0$ time-translates all states in $\{|n\rangle\}$ so that
they remain contained within it. States in which
$a_R^\star({\bf k})$ operators act on a state $|n\rangle$,
such as $a_R^\star({\bf q}_1)\cdots a_R^\star({\bf
q}_i)a_Q^\star({\bf k}_1)\cdots a_Q^\star({\bf
k}_j)|N\rangle$, are included in $\{|h\rangle\}$, but
excluded from $\{|n\rangle\}$. Such states are not
probabilistically interpretable.

As in all other gauge theories, Gauss's law is not an
equation of motion in CS theory. The operator ${\cal G}({\bf
x})$ used to implement Gauss's law is \begin{equation}
{\cal G}({\bf x}) = j_0({\bf x}) -\case 1/2
m\epsilon_{ij}F_{ij}({\bf x}),
\label{eq:gausslawop}
\end{equation}
and whereas $\partial_0 G={\cal G}$, $\partial_0\partial_0 G
= \partial_0 {\cal G} = 0$ is the equation of motion that
governs the behavior of this model. Further measures must be
taken to implement ${\cal G}=0$. We can conveniently express
${\cal G}$ in the form
\FL
\begin{equation}
{\cal G}({\bf x}) = m^{3/2}\sum{}[a_Q({\bf k}) +
a^\star_Q(-{\bf k})+\frac{j_0({\bf k})}{m^{3/2}}]e^{i{\bf k
\cdot x}},
\end{equation}
where $j_0({\bf k}) = \int d{\bf x}\,j_0({\bf x})e^{-i{\bf k
\cdot x}}$. We can define an operator $\Omega({\bf k})$ as
\begin{equation}
\Omega({\bf k}) = a_Q({\bf k}) + ({2m^{3/2}})^{-1}j_0({\bf
k}),
\end{equation}
so that
\begin{equation}
{\cal G}({\bf x}) = m^{3/2}\sum{}[\Omega({\bf k})e^{i{\bf k
\cdot x}} + \Omega^\star({\bf k})e^{-i{\bf k \cdot x}}].
\label{eq:GaussOmega}
\end{equation}
Similarly, we can write $A_0({\bf x})$ as
\FL
\begin{equation}
A_0({\bf x}) = im^{-1/2}\sum{}[\Omega({\bf k})e^{i{\bf k
\cdot x}} - \Omega^\star({\bf k})e^{-i{\bf k \cdot x}}].
\label{eq:AoOmega}
\end{equation}
We can therefore implement Gauss's law and the gauge
condition by embedding the theory in a subspace
$\{|\nu\rangle\}$ of another Hilbert space. The subspace
$\{|\nu\rangle\}$ consists of the states $|\nu\rangle$ which
satisfy the condition
\begin{equation}
\Omega({\bf k})|\nu\rangle = 0.
\label{eq:Omeganu}
\end{equation}
It can be easily seen from Eqs.~(\ref{eq:GaussOmega}) and
(\ref{eq:AoOmega}) that, in the physical subspace
$\{|\nu\rangle\}$, $\langle\nu^\prime|{\cal G}|\nu\rangle
=0$ and $\langle\nu^\prime|A_0|\nu\rangle =0$, so that both
Gauss's law and the gauge condition $A_0 = 0$ hold.
Moreover, the condition $\Omega({\bf k})|\nu\rangle = 0$,
once established, continues to hold at all other times
because
\begin{equation}
[H, \Omega({\bf k})] = 0
\label{eq:HOmega}
\end{equation}
so that $\Omega({\bf k})$ is an operator-valued constant.
This demonstrates that a state initially in the physical
subspace $\{|\nu\rangle\}$ will always remain entirely
contained within it as it develops under time evolution.

Consider now the unitary transformation $U=e^D$ where
\begin{equation}
D = -i\int{}d\,{\bf x}\,d{\bf y}\ \sum{}\frac{e^{i{\bf k
\cdot (x - y)}}}{k^2}\partial_i A_i({\bf x})j_0({\bf y}).
\label{eq:Dconfig}
\end{equation}
It is easy to show that
\begin{equation}
U^{-1}\Omega({\bf k})U = a_Q({\bf k}).
\end{equation}
We can use $U$ to establish a mapping that maps $\Omega({\bf
k}) \rightarrow a_Q({\bf k})$ and $\{|\nu\rangle\}
\rightarrow \{|n\rangle\}$, where $\{|n\rangle\}$ is the
subspace described previously in the paragraph following
Eq.~(\ref{eq:hip}). In this mapping, operators $P$ map into
$\tilde{P}$, i.e., $U^{-1}PU=\tilde{P}$. For example,
$\tilde{\Omega}({\bf k}) = a_Q({\bf k})$, and $\tilde{H} =
U^{-1}HU$ is given by
\begin{eqnarray}
\tilde{H} &=& H_0 - im^{-
1}\sum{}\frac{\epsilon_{ln}k_n}{k^2}j_l({\bf k})j_0(-{\bf
k})\nonumber\\
&-& i\sqrt{m}\sum{}\frac{\epsilon_{ij}k_j}{k^2}[a_Q({\bf
k})j_i(-{\bf k}) - a_Q^\star({\bf k})j_i({\bf k})].
\end{eqnarray}
If we expand $D$ in momentum space, we get $D = D_1 + D_2$
where
\FL
\begin{equation}
D_1 = ({2m^{3/2}})^{-1}\sum{}[a_R({\bf k})j_0(-{\bf k}) -
a_R^\star({\bf k})j_0({\bf k})]
\label{eq:D1}
\end{equation}
and
\begin{equation}
D_2 = i\sum{}\phi({\bf k})[a_Q({\bf k})j_0(-{\bf k}) +
a_Q^\star({\bf k})j_0({\bf k})].
\label{eq:D2}
\end{equation}
Since $D_2$ commutes with $a_Q({\bf k})$, it has no role in
transforming $\Omega({\bf k})$ into $a_Q({\bf k})$, and the
operator $V = e^{D_1}$ by itself achieves the same end as
$U$, i.e.,
\begin{equation}
V^{-1}\Omega({\bf k})V = a_Q({\bf k}).
\end{equation}
We can use $V$ to establish a second mapping of this theory,
in which operators map according to $P \rightarrow V^{-1}PV
= \hat{P}$. $\hat{\Omega}({\bf k}) = a_Q({\bf k})$, so that
$\hat{\Omega}$ and $\tilde{\Omega}$ are identical; under the
mapping $P \rightarrow V^{-1}PV = \hat{P}$, the subspace
$\{|\nu\rangle\}$ maps into the same subspace
$\{|n\rangle\}$ as under the mapping $P \rightarrow U^{-1}PU
= \tilde{P}$. But, in the case of other operators, $\hat{P}$
differs from $\tilde{P}$. For example, $\hat{H}$ is given by
\begin{eqnarray}
\hat{H} &=& H_0 - im^{-
1}\sum{}\frac{\epsilon_{ln}k_n}{k^2}j_l({\bf k})j_0(-{\bf
k})\nonumber\\
&-& i{m^{-3/2}}\sum{}\phi({\bf k})k_lj_l({\bf k})j_0(-{\bf
k})\nonumber\\
&-& i\sqrt{m}\sum{}\frac{\epsilon_{ij}k_j}{k^2}[a_Q({\bf
k})j_i(-{\bf k}) - a_Q^\star({\bf k})j_i({\bf
k})]\nonumber\\
&-& i\sum{}\phi({\bf k})k_i[a_Q({\bf k})j_i(-{\bf k}) -
a_Q^\star({\bf k})j_i({\bf k})]
\end{eqnarray}
Similarly, $\tilde{\psi}$ and $\hat{\psi}$ differ from each
other, although both project, from the correspondingly
defined vacuum states, electron states that implement
Gauss's law. $\tilde{\psi}$ and $\hat{\psi}$ are given by
$\tilde{\psi}({\bf x})=\exp[{\cal D}_{\mbox{\rm\scriptsize
U}}({\bf x})]\psi({\bf x})$ and by $\hat{\psi}({\bf x})
=\exp[{\cal D}_{\mbox{\rm\scriptsize V}}({\bf x})]\psi({\bf
x})$, \cite{comm} where
\begin{equation}
{\cal D}_{\mbox{\rm\scriptsize U}}({\bf x}) = -ie\int d{\bf
y}\sum{}\frac{e^{i{\bf k \cdot (x-
y)}}}{k^2}\partial_iA_i({\bf y})
\end{equation}
and
\begin{eqnarray}
{\cal D}_{\mbox{\rm\scriptsize V}}({\bf x}) &=& -ie\int
d{\bf y}\sum{}\frac{e^{i{\bf k \cdot (x-
y)}}}{k^2}\times\nonumber\\
&&\left[\partial_iA_i({\bf y}) + \frac{\phi({\bf
k})}{\sqrt{m}}k^2\epsilon_{ij}\partial_iA_j({\bf y})\right].
\end{eqnarray}
In the unitarily transformed representation, $a_Q({\bf
k})|n\rangle = 0$ is the form taken by the constraint that
implements Gauss's law and the gauge condition, when either
$U$ or $V$ is used to carry out the transformation.
$\tilde{J}$ and $\hat{J}$ are the forms into which the
Noether angular momentum operator $J$ is mapped when it is
unitarily transformed by $U$ and $V$, respectively. Both
these forms, $\tilde{J}$ and $\hat{J}$, are therefore
significant for the rotation of states in $\{|n\rangle\}$,
and it is of particular importance to observe that
$\tilde{J}$ and $\hat{J}$ differ from each other. $J$ is
given by
\begin{equation}
J = J_{\mbox{\scriptsize\rm g}} + J_{\mbox{\rm\scriptsize
e}},
\end{equation}
where $J_{\mbox{\scriptsize\rm g}}$ and
$J_{\mbox{\rm\scriptsize e}}$ are the angular momenta of the
gauge field and the spinors, respectively.
$J_{\mbox{\scriptsize\rm g}}$ and $J_{\mbox{\rm\scriptsize
e}}$ are given by
\FL
\begin{equation}
J_{\mbox{\scriptsize\rm g}} = -\int d{\bf
x}\,\epsilon_{ln}(\Pi_i x_l\partial_nA_i - G
x_l\partial_nA_0 + \Pi_l A_n)
\end{equation}
and
\begin{equation}
J_{\mbox{\rm\scriptsize e}} = -\int d{\bf x}\ (i\psi^\dagger
x_l\epsilon_{ln}\partial_n\psi+\case 1/2
\psi^\dagger{\gamma_0}\psi).
\label{eq:Je}
\end{equation}
Under the transformation mediated by $U$, $J \rightarrow
\tilde{J}$, and $\tilde{J} = J$, so that $J$ remains
untransformed. But, under the transformation mediated by
$V$, $J \rightarrow \hat{J}$ where $\hat{J} = J + {\cal J}$
and
\begin{eqnarray}
{\cal J} &=& - \sum{}\epsilon_{ln}k_l\frac{\partial\phi({\bf
k})}{\partial k_n}[a_Q^\star({\bf k})j_0({\bf k}) + a_Q({\bf
k})j_0(-{\bf k})]\nonumber\\
&+& {(2m)^{-
3/2}}\sum{}\epsilon_{ln}k_l\frac{\partial\phi({\bf
k})}{\partial k_n}j_0({\bf k})j_0(-{\bf k}).
\end{eqnarray}
We can support the preceding demonstration that $J$
transforms into itself under the unitary transformation
mediated by $U$, whereas it transforms into $J + {\cal J}$
under the unitary transformation mediated by $V$, with the
following observation: $D$ is an integral over operators and
functions which all transform as scalars under spatial
rotations. Since $J$ is the generator of spatial rotations,
the commutator $[J,D]$ must vanish. $D_1$ is not such an
integral over scalars, and there is therefore no similar
requirement that $[J,D_1]$ vanishes.

Since $U$ and $V$ map $\Omega({\bf k})$ into $a_Q({\bf k})$
in idential ways, we can conclude that the implementation of
Gauss's law is not responsible for the fact that $J$ is
transformed into $J + {\cal J}$ when $V$ is used to effect
the mapping. In fact, we can use the Baker-Campbell-
Hausdorff relation to construct an operator $W =
e^{D^\prime}$, where
\begin{eqnarray}
D^\prime &=&  i{(2m)^{-3/2}}\sum{}\phi({\bf k}){j_0({\bf
k})j_0(-{\bf k})}\nonumber\\
&-& i\sum{}\phi({\bf k})[a_Q({\bf k})j_0(-{\bf k}) +
a_Q^\star({\bf k})j_0({\bf k})],
\label{eq:Dprime}
\end{eqnarray}
so that $V = UW$. $W$ has the same effect as $V$ on $J$,
i.e. we find that
\begin{equation}
W^{-1}JW = J + {\cal J},
\end{equation}
although $W$ leaves $\Omega({\bf k})$ and ${\cal G}({\bf
x})$ untransformed and does not play any role in
implementing Gauss's law. $\phi({\bf k})$ is arbitrary, and
if we choose to set $\phi({\bf k}) = 0$, $U$ and $V$ become
identical. But if we choose $\phi({\bf k}) =
\sqrt{m}[\delta(k)/k]\tan^{-1}(k_2/k_1)$, then ${\cal J}$
becomes ${\cal J} = Q^2/4\pi m$, and accounts for the well-
known anyonic phase in the rotation of charged states
through $2\pi$.

In comparing $\tilde{H}$ with $\hat{H}$, we note that they
differ by some terms that include $a_Q({\bf k})$ or
$a_Q^\star({\bf k})$ as factors. Since both $\tilde{H}$ and
$\hat{H}$ are entirely free of $a_R({\bf k})$ and
$a_R^\star({\bf k})$ operators, $a_Q({\bf k})$ and
$a_Q^\star({\bf k})$ commute with every other operator that
appears in $\tilde{H}$ or $\hat{H}$. The terms which include
$a_Q({\bf k})$ or $a_Q^\star({\bf k})$ as factors therefore
do not affect the time evolution of state vectors in the
part of the subspace $\{|n\rangle\}$ that describes
observable particles (i.e., electrons or positrons); they
can neither produce projections on physical states, nor can
they contribute internal loops to radiative corrections.
They have no effect whatsoever on the physical predictions
of the theory and if they are arbitrarily amputated from
$\tilde{H}$ or $\hat{H}$, none of the physical predictions
are affected. The only other difference between $\tilde{H}$
and $\hat{H}$ is a total time derivative in $\hat{H}$, which
can be expressed alternatively as $h = i[H_0,\chi]$, as
$h=i[H,\chi]$, as $h=i[\tilde{H},\chi]$, or as
$h=i[\hat{H},\chi]$ where
\begin{equation}
\chi = -({2m})^{-3/2}\sum{}\phi({\bf k}){j_0({\bf k})j_0(-
{\bf k})}.
\end{equation}
The presence of $h$ in $\hat{H}$ is equivalent to unitarily
transforming the Hamiltonian as shown by $\hat{H} =
e^{i\chi}\tilde{H}e^{-i\chi}$. In earlier work, we
demonstrated that two Hamiltonians that are unitarily
equivalent in that fashion give rise to identical $S$-matrix
elements under very general conditions. \cite{hl}

We observe from these results that CS theory does give rise
to anyonic as well as to normal states: some states that
obey Gauss's law and the gauge condition rotate like normal
fermions; others show the arbitrary phase anomaly when the
angular momentum operator is used to generate rotations in
the plane. However, contrary to what has been suggested by
other authors, \cite{semenoff,swanson} it is not the
implementation of Gauss's law that is responsible for the
development of anyonic properties. States can develop an
anyonic angular momentum anomaly as an incidental byproduct
of the process by which Gauss's law is implemented, but the
change in the rotational properties of the state is not an
inevitable consequence of the implementation of Gauss's law.
Moreover, in corraboration of a result obtained by other
means, \cite{hagen} we find that regardless of whether the
arbitrary rotational phase develops, the anticommutation
rule that governs the electron field operator remains
unchanged. And that observation applies equally to the free
Dirac field and to the gauge-invariant electron field that
projects electrons that obey Gauss's law. The ``normal'' and
the ``anyonic'' operators are unitarily equivalent and both
obey Fermi-Dirac statistics.

\begin{center} {\bf ACKNOWLEDGMENTS} \end{center}

This research was supported by the Department of Energy
under Grant No. DE-FG02-92ER40716.00.

\end{document}